\documentstyle[12pt,epsf]{article}
\title{Random Lattice QCD and chiral effective theories}

\author{ O.V. Pavlovsky \thanks{e-mail address:
ovp@goa.bog.msu.su }  \\
{\em
Institute for Theoretical Problems of Microphysics,} \\
{\em Moscow State University } \\ {\em Moscow, 119992, Russian
Federation.} }

\date{ \ \ \  }

\begin{document}

\maketitle

\begin{abstract}
Resent developments in the Random Matrix and Random Lattice Theories
give a possibility to find low-energy theorems for many physical
models in the Born-Infeld form \cite{IIB}. In our approach that
based on the Random Lattice regularization of QCD we try to used the
similar ideas in the low-energy baryon physics for finding of the
low-energy theory for the chiral fields in the strong-coupling
regime.
\end{abstract}

\vspace{1cm}

PACS number(s): 11.15.Ha, 12.38.Gc, 12.39.Fe, 11.15.Me, 11.30.Cp,
11.30.Rd, 12.39.Dc

\section{Introduction and motivation}

The derivation of effective chiral theories from QCD has a long
history. This task has the strong motivation which come from the
necessity of the constructing of the QCD motivated baryon state
model. From the DIS experiments  we know, that the baryon consists
of the charged constituents (so called constituent quarks), but the
experiments with the spin of proton showed that only twenty percents
of the baryon spin can be associated with the charged constituents
\cite{spin}! Therefore, the realistic model for the low-energy
baryon state should describe the quark constituents as well as the
chiral degrees of freedom of the meson cloud around them. But what
is a theory that describe this chiral degrees of freedom and how
such theory can be derived from the origin theory - from QCD?

There are many conceptions and methods were proposed for finding the
answer on these questions so far. The first essential contribution
in this theme was done in \cite{witch}. Using the method of the
large $N$ expansion,  a effective chiral theory was suggested in the
form of the series of the chiral invariants. The second order theory
of such type is the well-known Skyrme model \cite{skyrme} of the
low-energy baryon state - the phenomenological unified theory for
mesons and baryons where the baryon treads as a topological soliton
of non-linear chiral fields. Another interesting approach was
proposed in \cite{novozh} where the Skyrme model was derived from
the integration of the chiral anomaly.

All such methods play the very essential role in the particle
physics and give a appropriate description  of the chiral field
behavior with the low tensity. However for studying of the chiral
field near the confinement boundary, all series of chiral
perturbations must be analyzed. The Chiral Bag Model \cite{bag} is a
very good illustration of this problem. In this model a boundary
between chiral fields and  color fields is specified by hand. In
such approach the agreement with the experiment dates is quit good
but in spite of this fact it is not clear now what is a physical
mechanism of formation of such chiral bag.

Fortunately today Lattice QCD numerical experiments give us a many
interesting information about the behavior of the color fields
(quarks and gluons) in the strong coupling regime. These dates are
very essential for understanding of the low-energy baryon's physics
and this approach is the empirical (and theoretical) basis of the
the baryon string model \cite{barstring}. Against a background of
these facts it would looks very astonishing that Skyrme model gives
a good agreement with the experiment dates although this model does
not describe the color degrees of freedom at all. So the question
about unification of these paradigms looks very essential now and
the seeking of the way of such unification naturally to begin with
an analyzing of the chiral limit of QCD on the lattice.

\section{Why we need in the Random Lattice QCD}

The attempts to obtain a chiral effective lagrangian from lattice
QCD had been performed many times a long ago. Using of the
well-known Brezin\&Gross trick \cite{Brezin} it could be possible to
perform the link's matrix integration in strong coupling regime and
obtain the various first order chiral effective theories
\cite{forder}.

In spite of first great success this approach had not been very
popular and origin of this stems from the fact that the approaches
from \cite{forder} does not take the possible to obtain any
corrections to the first order results. The lattice regularization
breaks a rotational symmetry of the initial theory from the
continues rotation group to a discrete group of rotations on fixed
angles. And the lattice regularization approach gives the correct
results only such tensors which are invariant by respect to this
discrete group. In particular using the ordinary Hyper-Cubical (HC)
lattice one can obtains the only first order effective theory and
for corrections this method generates non-rotational (non-lorentz)
invariant terms. For generating of the high-order effective field
theories more symmetrical lattice must be considered.

The problem of the rotational symmetry broking on a lattice has
attracted a principal attention for a long time. It was shown
\cite{f4} that in 4 dimension so-called  Body Centered Hyper-Cubical
(BCHC) or F4 lattice has the largest discrete symmetry  group. (BCHC
consists from the all sites of the HC lattice together with centers
of its elementary cells.)  This property of the BCHC lattice gives a
possibility to obtain the next-to-leading (NL) correction to the
first order of the chiral perturbation theory \cite{rebbi}. Results
of the works \cite{rebbi} are very essential for our analysis and
its point out on the effectiveness of this method. Moreover, this
results are very interesting from phenomenological point of view
because, as is well known \cite{skyrme}, the NL corrections violate
the scale invariance of the prototype (first order) chiral theory
that lead to generation of the chiral topological solitons
(Skyrmions). the Next-Leading order chiral effective theory that was
done in the \cite{rebbi} is in agreement with our phenomenological
propositions about this \cite{Gasser} and in our work we will used
the methodological ideas from \cite{rebbi} in order to define the
behavior of chiral field near the confinement surface.

As one could see, in order to solve our problem, the Next-Leading
order corrections are not enough. This theory has no solutions that
look like chiral "bag". Moreover, as would be shown later, near the
confinement surface (near the source of chiral field), the influence
of high order corrections became large and large. But the BCHC
lattice method gives the NL corrections only, the further use of
this method for the defining of the high order terms leads to
generation of non-relativistic (non-rotational) invariant terms. It
means that we need more a symmetric lattice than the BCHC lattice.

Unfortunately, the lattice which would be more symmetric than BCHC
lattice can not be constructed in 4 dimension. Moreover any method
based on a lattice with the fixed positions of sets has artifacts
concerned with priority directions that correspond with the basis
vectors of the lattice. Finally just these artifacts lead to the
problems with the rotational (relativistic) invariance and on this
evidence the using of the BCHC lattice is only half measure. For
solving of our problem the modification of the initial conception of
the lattice regularization must be performed. We need to find the
conception of the lattice regularization that has no any priority
directions. Fortunately this conception is known for a long time and
is called the Random Lattice Approach \cite{RL}.

The ideas of the Random Lattice was proposed by Voronoi and Deloune
and  today this method is very widely used in the modern science and
technology. For the quantum field theory method was modified by
Christ, Friedberg and Lee \cite{RL}. In these articles have been
shown that in order to obtain the restoration of the Lorentz
(rotational) invariance, it is necessary to perform an average over
an ensemble of random lattices. As result one get the averaging over
all possible directions and it is intuitively clear that this
procedure leads to the disappearance of the artifacts connected this
violation of the group of the space rotation.

But how to perform such random discretization? This procedure has
the tree steps:

1) Draw $N$ sites $x_i$ at random in the volume $V$.

2) Associate with each $x_i$ a so-called Voronoi cell $c_i$
$$
c_i = \{ x | d(x,x_i) \leq d(x,x_j), \forall j \neq i  \}
$$
where $d(x,y)$ is a distance between points $x$ and $y$. It is
means  that Voronoi cell $c_i$ consists of all points $x$ that are
closer to the center site $x_i$ than any other site.

3) Constrict the dual Delaune lattice by linking the center sites
of all Voronoi cells which share a common face.

After this if one consider the the big ensemble of such
Voronoi-Deloune random lattices based on  various distributions of
sites $x_i$, it possible to prove the origin rotational symmetry
will restored \cite{RL}. In our work we use this procedure for
obtaining of an effective chiral lagrangian from lattice QCD. This
methodological point of view this is a modification of the method
proposed in \cite{rebbi} on the case of the Random Lattice approach.

\section{From Lattice QCD to chiral lagrangians: step by step}

Now let me briefly remand a general steps of the algorithm of the
chiral lagrangian derivation from the lattice QCD that was proposed
in \cite{rebbi}. The starting point of our analysis is a standard
lattice action with Willson fermions
$$
Z= \int [DG] [D \bar{\psi} ] [D \psi ] \exp \{-S_{\mbox{pl}}(G)
-S_q(G,\bar{\psi},\psi) - S_J \}
$$
where:

1) plaquette gauge field term is
$$
S_{\mbox{pl}}= \frac{2N_{c}}{g^2} \sum_{pl} \left[1 - \frac{1}{N_c}
Re G_{x,\mu} G_{x-\mu,\nu} G^=_{x+\nu,\mu} G^+_{x,\nu} \right], \,
\, G_{x,\mu}=\exp\{ig\int_{\mbox{link}} dx'_\mu {\cal{A}}_\mu ( x' )
\};
$$

2)link fermions term is
$$
S_q = \sum_{x,\mu} Tr (\bar{A}_\mu (x) G_\mu (x) + G^+_\mu (x) A_\mu
(x) )
$$
$$
A_\mu(x)^a_b = \bar{\psi}_b(x+\mu) P^+_\mu \psi^a (x), \, \, \,
\bar{A}_\mu(x)^a_b = \bar{\psi}_b(x) P^-_\mu \psi^a (x+\mu)
$$
and $P^\pm_\mu=\frac{1}{2}(r \pm \gamma_\mu)$;

3) source term is
$$
S_J= \sum_x J^\alpha_\beta (x) M^\beta_\alpha (x), \, \, \,
M^\beta_\alpha = \frac{1}{N_c} \psi^{a,\beta}(x) \bar{\psi}_{a,
\alpha} (x).
$$

In order to realize the strong-coupling regime on the lattice let us
consider the limit of the large coupling constant $g$ ($g \to
\infty$). This limit was very wide studied \cite{forder} and main
result is that in such limit integral over the gauge field can be
performed. (Of course, the direct  integration is difficult since
the exists the plaquette term $S_{\mbox{pl}}$, but due to the
strong-coupling limit on the first step it possible to neglect such
plaquette contributions by respect to the contribution from link
integral $S_{q}$.  The plaquette contributions could be considered
in the systematic manner as perturbations by $1/g$ \cite{forder}.)

Let us consider the leading order contribution in this
strong-coupling expansion. The integrals over the gauge degrees of
freedom can be calculated into the large $N$ limit by using of the
standard procedure \cite{forder} and result of these calculations is
following
\begin{equation}
Z= \int [D \bar{\psi} ] [D \psi ] \exp \{-N \sum_{x,\nu} \mbox{Tr}
[F(\lambda(x,\nu))] - S_J \} \label{z1}
\end{equation}
where $\lambda_\nu = - M(x)P^-_\nu M(x+\nu) P^+_\nu $ and
$$
F(\lambda) = \mbox{Tr} [ (1 - \sqrt{1-\lambda}) ] - \mbox{Tr} [
\log(1 - \frac{1}{2}\sqrt{1-\lambda})  ]
$$

Now it would be very interesting to point out that the function
$F(\lambda)$ has the typical form of the Born-Infeld action with
first logarithmic correction. It is no casual fact. In \cite{IIB},
it was shown by means of the very similar technics that the
low-energy theory of the IIB superstring has a Born-Infeld form.
From the methodological point of view we perform a similar analysis
for QCD on the lattice and it would significant to note ahead of the
process of our proving that our result would has a Born-Infeld form
too.

Our next step is the integration over the fermion degrees of freedom
in (\ref{z1}). Using the source technics it was shown \cite{rebbi}
that integral (\ref{z1}) can be re-written into the form of the
integral over the unitary bosons matrix $M_x$
\begin{equation}
 Z=\int D M \exp S_{\mbox{eff}}(M).
 \label{z2}
\end{equation}

As a matter of principle, we already perform the transformation from
the color lattice degrees of freedom ($G$ and $\psi$) to the boson
lattice degrees of freedom ($M$). Now our task is to realize the
continuum limit of expression (\ref{z2}).

The nest step of our analysis correspond with the studying of the
stationary points of the lattice action $S_{\mbox{eff}}$.
Fortunately this is very well studied task \cite{gw}. This problem
is connected with well-known investigations of the critical behavior
of the chiral field on the lattice and with the problem of the phase
transformation on the lattice (for references see the issue
\cite{Rossi}). In \cite{rebbi}, it was shown that for our task the
stationary point is
$$
\hat{M}_0 = u_0 \hat{I}, \, \, u_0 (m_q=0, r=1)  = 1/4.
$$
Now one can expressed $M(x)$ in terms of the pseudoscalar Goldstone
bosons
$$
M=u_0 \exp( i \pi_i \tau_i \gamma_5/f_\pi )= u_0 [ U(x)
\frac{1+\gamma_5}{2} + U^{+}(x) \frac{1-\gamma_5}{2}]
$$
and the effective action is given in the form of the Taylor
expansion around this stationary point
\begin{equation}
S_{\mbox{eff}} (U) = - N \sum_{k=1}^\infty
\frac{F^{(k)}(\lambda_0)}{k!} \sum_{x,\nu} \mbox{Tr} [
(\lambda_\nu(x) - \lambda_0)^k ] \label{seff}
\end{equation}

Let us consider the expansion of the chiral field $U=\exp(i \pi_i
\tau_i /f_\pi)$ on the lattice around point $x$ (by respect to the
small step of the lattice $a$)
$$
U(x+\nu)= U(x) + a (\partial_\nu U(x) ) +
\frac{a^2}{2}(\partial^2_\nu U(x) ) + \cdots
$$
And for components of the Taylor expansion (\ref{seff}) one obtain
\begin{equation}
\begin{array}{ccrccc}
\mbox{Tr}[(\lambda_\nu(x) - \lambda_0)]  &=& -2 \lambda_0 \mbox{Tr}(\alpha)& & & \\
\mbox{Tr}[(\lambda_\nu(x) - \lambda_0)^2]&=&  2 \lambda^2_0 \mbox{Tr}(\alpha^2) & - 4 \lambda^2_0 \mbox{Tr}(\alpha)  & &  \\
\mbox{Tr}[(\lambda_\nu(x) - \lambda_0)^3]&=& -2 \lambda^3_0 \mbox{Tr}(\alpha^3) & + 6 \lambda^3_0 \mbox{Tr}(\alpha^2) & &  \\
\mbox{Tr}[(\lambda_\nu(x) - \lambda_0)^4]&=&  2 \lambda^4_0
\mbox{Tr}(\alpha^4) & - 8 \lambda^4_0 \mbox{Tr}(\alpha^3)  &
+ 4 \lambda^4_0 \mbox{Tr}(\alpha^2)&  \\
\mbox{Tr}[(\lambda_\nu(x) - \lambda_0)^5]&=& -2 \lambda^5_0
\mbox{Tr}(\alpha^5)& \cdots\cdots & \cdots\cdots & \cdots\cdots
\end{array}
\label{comp}
\end{equation} where $\alpha = a^2 \partial_\nu U \partial_\nu U^+ +
O(a^4)$.

Expressions (\ref{comp}) are very essential because these are a
simplest illustration of all aspects of the violation of the
rotational symmetry on the lattice. For this moment we specially say
nothing about the structure of our lattice. We try to formulate our
result as general as possible and all information about the lattice
contains into the vectors $\nu$ that correspond with the basic
vectors of the lattice (for example, the vectors $\nu$ for the
Hyper-Cubical lattice are the Cartesian basic vectors $\vec i$,
$\vec j$, $\vec k$ and $\vec t$). The leading order part can be
calculated trivially. Indeed, using the simple Hyper-Cubical lattice
where $\nu=i,j \, :$ $i=(1 \ldots 4)$ it is easy to show that the
leading order contribution is the prototype chiral lagrangian
\begin{equation}
 P_{O(p^2)} \sim \mbox{tr} [ \partial_\mu U
\partial^\mu U^+ ]
\label{lorder}
\end{equation}

As I said before the rotational symmetry violation arguments does
not admit to use the HC lattice calculation for the next-leading
order contributions.  For obtaining of these contributions a more
symmetrical lattice must be used. In \cite{f4} it was shown that
this lattice is a Body Centered Hyper-Cubical (BCHC) or F4 lattice.
The basis vectors of this lattice are
\begin{equation}
\mbox{BCHL} \rightarrow  \nu=\nu^{\alpha}_{ij} = \frac{1}{\sqrt{2}}
(e_i+\alpha e_j), (1\leq i < j \leq 4, \alpha \pm 1)
\label{bas1}
\end{equation}
and one can show that the next-leading order contribution
(\ref{bas1}) in this basis is following
$$
P_{O(p^4)} \sim \mbox{tr}[(\partial_\mu U \partial^\mu U^+)^2] +
\frac{1}{2} \mbox{tr}[\partial_\mu U \partial^\nu U^+
\partial_\mu U \partial^\nu U^+]
$$

It is easy to see that this is just we expect to receive because
this contribution was obtained from many another approaches
\cite{witch,novozh}. But unfortunately, this method can not directly
used for finding next contributions and the origin of this fact is
again the violation of the rotation symmetry  but now on the  F4
lattice. Moreover, there are no any more symmetrical lattice with
fixed position of sites in 4-dimensional \cite{f4}. It means that we
are need in the absolutely different lattice conception that
guaranteed the restoration of the initial symmetries. Fortunately
this conception is known now. This is a Random Lattice conception
(RL) \cite{RL}.

The basic idea of the RL is the averaging over the big ensemble of
various lattices with random distributions of sites and it possible
to show that such averaging leads to the restoration of the
rotational invariance. There are two methods of the realization of
such scheme. A first one based on the Christ, Friedberg and Lee
(CFL) technics. The basis of vectors in the CFL method is following
$$
\mbox{RL} \rightarrow  \nu=\nu^{\mu}_{ij} = e^{\mu}_{ij}
s_{ij}/l_{ij}
$$
where $s_{ij}$ is a volume of the corresponding 3-dimensional
boundary surface of the Voronoi cells  and $l_{ij}=| \vec r_i - \vec
r_j|$ is the length of link. Using the summation formulas from
\cite{RL} one get that after the averaging only pairs are survive
\begin{equation}
< \prod_a (e^a)_\nu > = \sum_{\mbox{pairings}} \,
\prod_{\mbox{pairs}} < (e^a)_i (e^b)_i> \label{pair}
\end{equation}

At other hand, the result (\ref{pair}) could be obtained by means of
the following trick \cite{david, bogacz}. For beginning let us
consider a lattice with fixed position (for simplicity it possible
to use the trivial HC lattice) in a flat space. Now let us consider
small deformations of the geometry of this space ($\gamma_{ij} \to
g_{ij}$). Using of this idea one can rewrite the problem of the
random lattice averaging in the terms of the random surface
\cite{david}. This is the standard quantum gravity task and using
the methods of the Matrix Theory one can show that our result
(\ref{pair}) is just the direct consequence of well-known Wick's
Theorem about the pairings \cite{RM}.

The expression (\ref{pair}) gives us possibility to calculate all
terms in expansion (\ref{comp}). Let us consider just the first
column in the expression (\ref{comp}). It is easy to show that
either of these is proportional to some power of the leading order
contribution (\ref{lorder})
\begin{equation}
\begin{array}{ccccc}
\mbox{Tr}[\alpha] &\sim& \mbox{tr} [ \partial_\mu U \partial^\mu U^+ ]& + & \cdots \\
\mbox{Tr}[\alpha^2] &\sim& \mbox{tr} [ (\partial_\mu U \partial^\mu U^+)^2 ]& + & \cdots \\
\mbox{Tr}[\alpha^3] &\sim& \mbox{tr} [ (\partial_\mu U \partial^\mu U^+)^3 ]& + & \cdots \\
\cdots\cdots & \cdots & \cdots\cdots & \cdots & \cdots\cdots \\
\mbox{Tr}[\alpha^n] &\sim& \mbox{tr} [ (\partial_\mu U \partial^\mu
U^+)^n ]& + &
\cdots\\
\cdots\cdots & \cdots & \cdots\cdots & \cdots & \cdots\cdots
\end{array}
\label{ser}
\end{equation}
Substituting (\ref{ser}) into the (\ref{seff}) and collecting of all
terms which depend on the power of the prototype lagrangian one
obtain following expression for the effective chiral lagrangian
\begin{equation}
{\cal L}_{\mbox{eff}} \sim - \mbox{tr} \left[ 1 - \sqrt{1-1/\beta^2
\partial_\mu U \partial^\mu U^+ } \right] - \mbox{tr} \left[ \log(1 -
\frac{1}{2}(1-  \sqrt{1-1/\beta^2
\partial_\mu U \partial^\mu U^+ })) \right] + \cdots
\label{otvet}
\end{equation}
where $\cdots$ are all another terms (in particular the Skyrme term)
and $\beta$ is a effective coupling constant that depend on the
value of our stationary point $u_0$.

Now let discussed the result (\ref{otvet}). It was obtained that the
some part of chiral effective action has a Born-Infeld form plus
first logarithmic correction to it. In \cite{soliton}, it was shown
that such form of the effective action play a very essential role in
the problem of the chiral bag formation because just these
square-root terms generate the step-like distribution solutions that
can be interpreted as internal phase into the two-phase model of the
low-energy baryon states. Another terms play essential role only on
large distance from the confinement surface and can be considered as
corrections.

\section{Conclusions}
The aim of this paper is to derive the chiral effective lagrangian
from QCD on the lattice at the strong coupling limit. We find that
this theory looks like a Born-Infeld theory for the prototype chiral
lagrangian. Such form of the effective lagrangian is expected. From
the methodological point of view our consideration is very similar
with the low-energy theorem in string theory that lead to the
Born-Infeld action \cite{IIB}. Moreover, in \cite{soliton}, it was
shown that Chiral Born-Infeld Theory (without logarithmic
corrections) has very interesting   "bag"-like solution for chiral
fields. It was additional motivation of our work.

The Chiral Born-Infeld theory is a good candidate on the role of the
effective chiral theory and the model for the chiral cloud of the
baryons. In this model one can find not only spherical "bags", it is
possible to also study the "string"-like, toroidal or
"\textbf{Y}-Sign"-like solutions, or some another geometry. The
geometry of the confinement surface  depends directly on the model
of color confinement and it would be very interesting to use, for
example, the Lattice QCD simulations for the color degrees of
freedom in combination with our model for the external chiral field.

\vspace{1cm}

This work is partially supported by the Russian Federation
President's Grant 1450-2003-2. The hospitality and financial support
of the ICT* in Trento is gratefully acknowledged.

\end{document}